\documentclass[a4paper,12pt]{article}
\usepackage{cite}
\newcommand{\be}{\begin{equation}}
\newcommand{\ee}{\end{equation}}
\newcommand{\bea}{\begin{eqnarray}}
\newcommand{\nn}{\nonumber}
\newcommand{\eea}{\end{eqnarray}}

\begin{document}

\begin{titlepage}

\begin{flushright}
UB-ECM-PF-04/03\\
\end{flushright}
\begin{centering}
\vspace{.3in}
{\large{\bf Area Spectrum of Kerr and extremal Kerr Black Holes\\ from Quasinormal Modes}}
\\

\vspace{.5in} {\bf Mohammad R. Setare$^{1}$ and Elias C.
Vagenas$^{2}$ }\\
\vspace{.3in} $^{1}$\,Physics Dept., Inst. for Studies in Theo. Physics and
Mathematics (IPM)\\
P.O. Box 19395-5531, Tehran, Iran\\
rezakord@ipm.ir\\
\vspace{0.4in}

$^{2}$\, Departament d'Estructura i Constituents de la Mat\`{e}ria\\
and\\ CER for Astrophysics, Particle Physics and Cosmology\\
Universitat de Barcelona\\
Av. Diagonal 647, E-08028 Barcelona\\
Spain\\
evagenas@ecm.ub.es\\
\end{centering}

\vspace{0.7in}

\begin{abstract}
Motivated by the recent interest in quantization of black hole area spectrum, we consider the area spectrum of
Kerr and extremal Kerr black holes. Based on the proposal by Bekenstein and others that the black hole area
spectrum is discrete and equally spaced, we implement Kunstatter's method to derive the area spectrum for the Kerr
and extremal Kerr black holes. The real part of the quasinormal frequencies of Kerr black hole used for this
computation is of the form $m\Omega$ where $\Omega$ is the angular velocity of the black hole horizon. The
resulting spectrum is discrete but not as expected uniformly spaced. Thus, we infer that the function describing
the real part of quasinormal frequencies of Kerr black hole is not the correct one. This conclusion is in
agreement with the numerical results for the highly damped quasinormal modes of Kerr black hole recently presented
by Berti, Cardoso and Yoshida. On the contrary, extremal Kerr black hole is shown to have a discrete area spectrum
which in addition is evenly spaced. The area spacing derived in our analysis for the extremal Kerr black hole area
spectrum is not proportional to $\ln 3$. Therefore, it does not  give support to Hod's statement that the area
spectrum $A_{n}=(4l^{2}_{p}ln 3)n$ should be valid for a generic Kerr-Newman black hole.
\end{abstract}


\end{titlepage}

\newpage
\baselineskip=18pt

\section{Introduction}
Hawking was the first to observe that black holes are not completely black but they emit radiation \cite{haw1}.
This radiation is essentially thermal and the black hole emits field quanta of all frequencies. Now, may be one
can ask this question: do quantum black holes have a discrete spectrum? Several authors have raised the
possibility that Hawking radiation might in fact have a discrete spectrum. This idea was first proposed by
Bekenstein in 1974 \cite{bek1}. According to Bekenstein's proposal, the eigenvalues of the black hole event
horizon area are of the form
\be
A_n=\alpha  l_{p}^{2} n
\hspace{1ex},
\ee
where $\alpha$ is a dimensionless
constant, $n$ ranges over positive integers and, by using gravitational units $G=c=1$, $l_p=\hbar ^{1/2}$ is the
Planck length. Endorsement for Bekenstein's proposal was provided by the observation that the area of the horizon
$A$ behaves, for a slowly changing black hole, as an adiabatic invariant \cite {bek2}. It is significant that a
classical adiabatic invariant corresponds to a quantum observable with a discrete spectrum, by virtue of
Ehrenfest's principle. Moreover, the possibility of a connection between the quasinormal frequencies of black
holes and the quantum properties of the entropy spectrum was first observed by Bekenstein \cite{bek3}, and further
developed by Hod \cite{hod1}. Bekenstein noted that Bohr's correspondence principle implies that frequencies
characterizing transitions between energy levels of a quantum black hole at large quantum numbers correspond to
the black hole's classical oscillation (quasinormal) frequencies (see also \cite{{kok1},{nol}}). In particular,
Hod proposed that the real part of the quasinormal frequencies, in the infinite damping limit (i.e. the
$n\rightarrow\infty$ limit), might be related via Bohr's  correspondence principle to the fundamental quanta of
mass and angular momentum (see also \cite{bek4}--\hspace{-0.1ex}\cite{pad3}).

\par\noindent
The quasinormal modes (QNMs) of black holes are the characteristic ringing frequencies which result from their
perturbations \cite{chandra} and provide a unique signature of these objects \cite{kok1}, possible to be observed
in gravitational waves. In asymptotically flat spacetimes the idea of QNMs started with the work of Regge and
Wheeler \cite{reggeW} where the stability of a Schwarzschild black hole was tested, and were first numerically
computed by Chandrasekhar and Detweiler several years later \cite{Chandra1}. QNMs have nowadays motivated a flurry
of activity in different contexts: in AdS/CFT correspondence \cite{Horowitz-Hubeny}--\hspace{-0.1ex}\cite{carli},
when considering thermodynamic properties of black holes in loop quantum gravity
\cite{dry}--\hspace{-0.1ex}\cite{hon}, in the context of possible connection with critical collapse
\cite{Horowitz-Hubeny,BHCC,kim}.

\par\noindent
In the present paper  we extend directly the Kunstatter's approach \cite{kun} to determine mass and area spectrum
of Kerr and extreme Kerr black holes\footnote{Similar works on evaluating the area spectrum of other types of
black holes are \cite{set1,set2,set3}.}. According to this approach, an adiabatic invariant $I=\int {dE\over
\omega(E)}$, where $E$ is the energy of system and $\omega(E)$ is the vibrational frequency, has an equally spaced
spectrum, i.e., $I\approx n\hbar$, applying the Bohr-Sommerfeld quantization at the large $n$ limit. The reminder
of the paper is organized as follows. In Section 2 we implement Kunstatter's approach to a Kerr black hole.
Although, the area spectrum is found to be discrete, it is not equally spaced. In Section 3 we consider the area
spectrum of an extremal Kerr black hole. In this case, the spectrum is discrete and evenly spaced. Finally,
Section 4 is devoted to a brief summary of results and concluding remarks.

\section{Kerr Black Hole}
The metric of a four-dimensional Kerr black hole given in Boyer-Lindquist coordinates is
\be
ds^{2}=-(1-\frac{2Mr}{\Sigma})dt^{2}-\frac{4Mar
\sin^{2}\theta}{\Sigma}dtd\varphi+\frac{\Sigma}{\Delta}dr^{2}+
\Sigma d\theta^{2}+(r^2+a^2+2Ma^2r\sin^{2}\theta)
\sin^{2}\theta d\varphi^{2}
\label{met}
\ee
where
\be
\Delta=r^2-2Mr+a^2
\label{deleq}
\ee
\be
\Sigma=r^2+a^2\cos^{\theta}
\label{sigeq}
\ee
and $M$ is the mass of black hole.
The roots of $\Delta$ are given by
\be
r_{\pm}=M\pm \sqrt{M^{2}-a^{2}}
\label{root}
\ee
where $r_{+}$ is the radius of the event (outer) black
hole horizon and $r_{-}$ is the radius of the inner black hole horizon.
In addition, we have defined the specific angular momentum as
\be
a=\frac{J}{M}
\label{aeq}
\ee
where $J$ is the angular momentum of the black hole. The
Kerr black hole is rotating with angular velocity
\bea
\Omega&=&\frac{a}{r_{+}^{2}+a^{2}}\\
&=&\frac{J}{2M\left(M^{2}+\sqrt{M^{4}-J^{2}}\right)}
\label{angve}
\eea
which has been evaluated on the event black hole horizon.
In gravitational units, the Kerr black hole horizon area
and its Hawking temperature are given, respectively, by
\bea
A&=&4\pi (r_{+}^{2}+a^{2})\\
&=&8\pi\left(M^{2}+\sqrt{M^{4}-J^{2}}\right)
\label{area}
\eea
and
\bea
T_{H}&=&\frac{r_{+}-r_{-}}{A}\\
&=&\frac{\sqrt{M^{4}-J^{2}}}{4\pi M\left(M^{2}+\sqrt{M^{4}-J^{2}}\right)}
\hspace{1ex}.
\label{temeq}
\eea
By applying Bohr's correspondence principle,
Hod\cite{hod2} conjectured that the real part of the asymptotic
quasinormal frequencies of Kerr black hole is given by the formula
\be
\omega_{R}=T_{H}\ln3+m\Omega
\label{quasi},
\ee
where $m$ is the azimuthal eigenvalue of the oscillation.
There was compelling evidence that the conjectured
formula (\ref{quasi}) must be wrong.

\par\noindent
A systematic exploration of the behavior of the first few overtones,
i.e., small values of the principal quantum number $n$, was first accomplished by Onozawa \cite{onoz}.
Onozawa used the Leaver's continued fraction method to carry out the numerical calculations.
Berti and Kokkotas \cite{berti1} confirmed Onozawa's results and extended them to higher overtones,
i.e., highly damped QNMs.
They found that the formula conjectured by Hod, i.e., equation (\ref{quasi}), does not seem to provide
a good fit to the asymptotic modes. Furthermore, Berti and Kokkotas showed that, as the mode order increases,
modes having their orbital angular momentum eigenvalue $l$ and azimuthal eigenvalue $m$ to satisfy
$l=m=2$, are fitted extremely well
by the relation\footnote{A more sophisticated study of highly damped Kerr QNMs is performed in \cite{berti2}.
In this work the authors provide complementing and clarifying results that were presented in
previous works\cite{onoz,berti1}.}
\be
\omega=2\Omega + \mathit{i} 2\pi T_{H} n
\hspace{1ex}.
\ee
where the term $2\pi T_{H}$ that appears in the imaginary part of the mode frequencies is the surface gravity
of the event horizon of the Kerr black hole.
Therefore, the real part of the asymptotic frequencies having $l=m=2$ is
given by the expression
\be
\omega_{R}=2\Omega
\hspace{1ex}.
\ee

\par\noindent
Hod studied again analytically the QNMs of Kerr black hole \cite{hod3}
and he concluded that the asymptotic quasinormal
frequencies of Kerr black hole are given by the simpler expression
\be
\omega_{R}=m\Omega
\label{quasi1}
\ee
which is obviously in agreement with the aforesaid numerical results of Berti and Kokkotas.
This classical frequency plays an important role in the dynamics of the black hole
and is relevant to its quantum properties \cite{{hod1},{dry}}.

\par\noindent
Given a system with energy $E$ and vibrational frequency $\omega(E)$, one can show that the
quantity
\be
I=\int \frac{dE}{\omega(E)}
\label{adiabatic}
\ee
where $dE=dM$, is an adiabatic invariant \cite{arnold}
and as already mentioned in the Introduction, via Bohr-Sommerfeld quantization has an equally spaced spectrum in
the large $n$ limit
\be
I \approx n\hbar \hspace{1ex}.
\label{smi}
\ee

\par\noindent
Bekenstein showed that for the case of black holes the adiabatic invariants, namely quantities
which vary very slowly compared to variations of the external perturbations on the black hole, are
the black hole horizon areas \cite{bek3,bek5}\footnote{For some examples see \cite{mayo}.}.

\par\noindent
Exploiting the idea of adiabatic invariants and the statement by Bekenstein,
Kunstatter \cite{kun} derived for the $d(\geq 4)$-dimensional Schwarzschild black holes an
equally spaced entropy spectrum.
A key point to Kunstatter's approach was that the first law of black hole thermodynamics
for the case of a Schwarzschild black hole is of the form
\be
dM=\frac{1}{4}T_{H}dA
\label{sch1law}
\ee
and thus, using (\ref{sch1law}), the adiabatic invariant for the Schwarzschild black hole takes the form
\be
I= \int \frac{dM}{\omega_{R}}
\hspace{1ex}.
\ee
This is of the form (\ref{adiabatic}) where $dM=dE$ and $\omega(E)$ has been replaced by the real part of
asymptotic QNMs of Schwarzschild black hole as proposed by Hod \cite{hod1} since $\omega_{R}\sim T_{H}$.

\par\noindent
We now extend Kunstatter's approach to the case of Kerr black hole.
The first law of black hole thermodynamics now takes the form
\be
dM=\frac{1}{4}T_{H}dA+\Omega dJ
\label{kerr1law}
\ee
and obviously the corresponding expression for  adiabatic invariant is now given by the expression
\be
I= \int \frac{dM-\Omega dJ}{\omega_{R}}
\label{adiabkerr}
\hspace{1ex}.
\ee
The real part ($\omega_{R}$) of the asymptotic (highly damped) quasinormal frequencies
of the Kerr black hole is given by equation (\ref{quasi1})
and the angular velocity is given by equation (\ref{angve}).

\par\noindent
Therefore, the adiabatically invariant integral
(\ref{adiabkerr}) is written as
\be
I=\frac{2}{mJ}\int M \left(M^{2}+\sqrt{M^{4}-J^{2}}\right)\,\, dM - \frac{1}{m}\int dJ
\label{integral}
\ee
and after integration, we get
\be
I=\frac{1}{2mJ}\left[ M^{2}\left( M^{2}+\sqrt{M^{4}-{J^2}}\right)- J^{2}
ln\left(M^{2}+\sqrt{M^{4}-{J^2}}\right)\right] -\frac{1}{m}J
\label{action1}
\hspace{1ex}.
\ee
By equating expressions (\ref{smi}) and (\ref{action1}), we get
\be
M^{2}\bar{A}-J^{2}ln\bar{A}-c =\left(2mJl^{2}_{p}\right)n
\label{nonalgebraic}
\ee
where  the parameter $c$ is equal to $2 J^{2}$, the quantity $\bar{A}$ is the reduced horizon area
\be
\bar{A}=\frac{A}{8\pi}
\ee
and the area $A$ of the Kerr black hole horizon is given by equation (\ref{area}).
The solution to equation (\ref{nonalgebraic}) is the principal
branch of Lambert W-function (see Appendix) and thus the area of the Kerr black hole is written
\be
A=8\pi\left(-\frac{J^{2}}{M^{2}}\right) W_{0}[z]
\label{kerrsoln1}
\ee
where the argument of Lambert W-function is
given by
\be
z=\left(-\frac{M^{2}}{J^{2}}\right)e^{-2\left(1+\frac{m\hbar}{J} n\right)} \hspace{1ex}.
\ee
Since we are only interested in highly damped quasinormal frequencies, i.e. $n\rightarrow\infty$,
we keep only the first term of the series expansion of the
Lambert W-function and we get
\be
A=8\pi e^{-2\left(1+\frac{m\hbar}{J} n\right)}
\label{Kerrsoln}
\ee
It is obvious that the area spectrum, although
discrete, is not equivalently spaced even to first order. Since the area spectrum of Kerr black hole has been
proven by Bekenstein \cite{bek1,bek3} and others \cite{makela,gour}, to be discrete and uniformly spaced, we
conclude that the function that we have used in our computation as the real part of the asymptotic
quasinormal frequencies of Kerr black hole, i.e. expression (\ref{quasi1}), is not the correct one.
Our conclusion which is based on an analytical computation, is in agreement with the very recent numerical
results of Berti, Cardoso and Yoshida\cite{berti} who computed the very highly damped
QNMs of Kerr black hole and showed that the
real part of the corresponding quasinormal frequencies are not given by simple polynomial
functions of the black hole temperature $T_{H}$ and angular velocity $\Omega$ (or their inverses).

\par\noindent
A couple of comments are in order. Firstly, a crucial feature for the present analysis is that the highly damped
quasinormal frequencies depend only on the black hole parameters and are independent of the characteristics of the
perturbation field. However, there are works where the quasinormal frequencies depend on the orbital angular
momentum, $l$, which is a characteristic of the perturbation field \cite{cardos,abd,set3,med1} \footnote{The
$l$-dependence of the real part of the frequency was also proven in the context of dirty black holes \cite{med2}. More
specifically, it was proven that in the ``squeezed-horizon'' limit the real part of the frequency of dirty black
holes goes almost linearly with the orbital angular momentum $l$.}. Secondly, an equivalently important feature is
the universality of the fundamental lower bound $(\Delta A)$, i.e. it is independent of the black hole parameters.
In the present analysis for the area spectrum of Kerr black hole, it is evident by checking expression
(\ref{kerrsoln1}), that the area spectrum depends on both black hole parameters, i.e. the mass $M$ and the angular
momentum $J$.

\par\noindent
Finally, it is noteworthy that if one selects as a function for the real part of the quasinormal frequencies, not
the second but the first term of equation (\ref{quasi}), then the area spectrum of Kerr black hole will be
discrete and uniformly spaced. More precisely, in this case the area spectrum of Kerr black hole is given as
\be
A_{n}=\left(\Delta A\right)n
\label{conj1}
\ee
where the area spacing reads
\be
\left(\Delta A\right) =4
l^{2}_{p}\ln 3
\label{conj2}
\hspace{1ex}.
\ee
Although this result is exactly what we would like to
derive\footnote{Based on the universality of the black hole entropy and the universality of the lower bound, Hod
had conjectured  that expressions (\ref{conj1}) and (\ref{conj2}) should be valid for a generic Kerr-Newman black
hole \cite{hod1}.}, at the moment it is just a computational coincidence and it should not be seriously
considered. Furthermore, as we have already mentioned before, Kunstatter \cite{kun} had proven
that the adiabatic invariant $I$ for the case of
D-dimensional Schwarzschild black hole is
\be
I\propto\int\frac{dE}{T_{H}}
\ee
but he also raised the question whether this relationship also holds for charged or rotating black holes.


\section{Extremal Kerr Black Hole}
In order to evaluate the area spectrum for the extremal Kerr black hole we have to adopt a new prescription
since the horizon area is not an adiabatic invariant for the case of extremal Kerr black hole \cite{bek3,bek5}.
We therefore follow the Kunstatter's approach as presented in the previous section.
The action is  written as before
\be
I=\int \frac{dE}{\omega_{R}}
\label{action}
\ee
where $dE=dM$, but the real part of the quasinormal frequency is now given as
\be
\omega_{R}=
\frac{m}{2M}
\label{extremalqnf}
\ee
which is easily derived by imposing the extremality condition, i.e., $J^{2}=M^{4}$,
upon equation (\ref{quasi1}), that we rewrite here
\bea
\omega_{R}&=&m\Omega\nn\\
&=&m\frac{J}{2M\left(M^{2}+\sqrt{M^{4}-J^{2}}\right)}
\hspace{1ex}.
\eea
Therefore, the integral (\ref{action}) yields
\be
I=\frac{1}{m}M^{2}
\label{action3}
\ee
and by equating expressions (\ref{action3}) and (\ref{smi}), we get
\be M^{2}=ml^{2}_{p}\,n
\label{extremalmass}
\label{1ex}.
\ee
In the context of black hole thermodynamics, the area of the extremal Kerr
black hole horizon is given by
\be
A=8\pi M^{2}
\ee
and thus by substituting (\ref{extremalmass}), we get
\be
A_{n}=8\pi ml^{2}_{p}n
\label{areaextrkerr}
\hspace{1ex}.
\ee
It is obvious that the area spectrum of extremal Kerr black hole is equally spaced and discrete.
By setting $m=1$, the area spectrum (\ref{areaextrkerr}) is
exactly the same with the one derived for the case of quasi-extreme Kerr black hole in \cite{abd}.
Furthermore, the universal (i.e. independent of black hole parameters) lower bound
that we derived here, i.e.,  $(\Delta A)=8\pi l^{2}_{p}$,
was also derived by Bekenstein but it was valid only for the non-extremal Kerr-Newman black hole.
Bekenstein's analysis failed for the the case of extremal Kerr-Newman black hole and so one could not deduce that
its area eigenvalues were evenly spaced \cite{bek1,bek2}. Therefore, by employing the Bekenstein-Hawking area
formula for the black hole entropy, i.e.,
\be
S_{BH}=\frac{1}{4l^{2}_{p}}A
\hspace{1ex},
\ee
the entropy of extremal Kerr black hole is now given by
\be
S_{n}=\left(2\pi m\right) n
\hspace{1ex}.
\ee

\section{Conclusions}
In this paper we have evaluated analytically the area spectrums of Kerr and extremal Kerr black holes by
implementing Kunstatter's approach. The area spectrum of Kerr black hole was derived by using as real part of its
quasinormal frequencies a function of the form $m\Omega$. It was shown that the area spectrum is discrete but not
evenly spaced. Furthermore, an unexpected feature of the area spectrum is that it depends explicitly on the Kerr
black hole parameters, i.e. the mass and the angular momentum. It is clear that since the novel numerical results
show that the real part of the quasinormal frequencies of Kerr black hole is not just a simple polynomial function
of its Hawking temperature and its angular velocity (or their inverses), further theoretical\hspace{1ex}study is
needed.

We have also shown that the area spectrum of extremal Kerr black hole is discrete and equidistant. The
corresponding lower bound is universal, i.e. independent of the black hole parameters, but it is not proportional
to $\ln3$. Therefore, it does not provide any support to Hod's statement that the area spectrum of the form
$A_{n}=(4l^{2}_{p}\ln3)n$ should be valid for a generic Kerr-Newman black hole.

Finally, it  is now known that the asymptotic quasinormal frequencies of Reissner-Nordstr\"om black hole are given
by a QNM condition involving exponentials of its temperature. It seems likely that the asymptotic quasinormal
frequencies of Kerr black hole will also be described by such an analytic formula. We hope to return to this issue
in a future work.


\section{Acknowledgments}
The authors wish to thank Kostas Kokkotas for useful correspondence. The work of E.C.V. has been supported by the
European Research and Training Network ``EUROGRID-Discrete Random Geometries: from Solid State Physics to Quantum
Gravity" (HPRN-CT-1999-00161).


\section{Appendix}
Consideration of Lambert W function can be traced back to J. Lambert around 1758, and later,
it was considered by L. Euler but it was recently established
as a special function of mathematics on its own
\cite{lambert1}\footnote{This publication in its
technical report form appeared in 1993.}.
\par\noindent
The Lambert W function is defined to be the function satisfying
\be
W[z]e^{W[z]}=z
\hspace{1ex}.
\ee
It is a multivalued function defined in general for $z$ complex and assuming values $W[z]$ complex.
If $z$ is real and $z<-1/e$, then $W[z]$ is multivalued complex.
If $z$ is real and $-1/e \leq z \leq 0$, there are two possible real values of $W[z]$.
The one real value of $W[z]$ is the branch satisfying $-1\leq W[z]$, denoted by $W_{0}[z]$,
and it is called the principal branch of the W function. The other branch is $W[z] \leq -1$
and is denoted by $W_{-1}[z]$. If $z$ is real and $z\geq 0$, there is a single real value for $W[z]$
which also belongs to the principal branch $W_{0}[z]$. Special values of the principal branch of
the Lambert W function are $W_{0}[0]=0$ and $W_{0}[-1/e]=-1$.
\par\noindent
The Taylor series of $W_0[z]$ about $z=0$ can be found using the Lagrange
inversion theorem and is given by \cite{lambert2}
\be
W[z]=\sum^{\infty}_{n=1}\frac{(-1)^{n-1}n^{n-2}}{(n-1)!}z^{n}=
z-z^{2}+\frac{3}{2}z^{3}-\frac{8}{3}z^{4}+\frac{125}{24}z^{5}-\frac{54}{5}z^{6}+\dots\nn
\hspace{1ex}.
\ee
The ratio test establishes that this series converges if $|z|<1/e$.


\end{document}